\documentclass[referee]{raa}            

\usepackage{graphicx,times}             
\input{epsf.sty}                        
\input{psfig.sty}                       

\begin{document}

   \title{Diagnostics  From Three Rising Submillimeter Bursts  
}

   \volnopage{Vol.0 (200x) No.0, 000--000}      
   \setcounter{page}{1}          

\author{A. H. Zhou
	\inst{1,2}
	\and J. P. Li
	\inst{1,2}
	\and X. D. Wang
	\inst{3}
}

\institute{Key Laboratory of Dark Matter and Space Astronomy, Chinese Academy of Sciences; {\it zhouah@pmo.ac.cn}\\
	\and
	Purple Mountain Observatory, Chinese Academy of Sciences, Nanjing 210008\\
	\and
	Hohai University, Nanjing 210098, China\\
}

   \date{Received~~2015 month day; accepted~~2015~~month day}

\abstract{In the paper we investigate three novel rising submillimeter (THz) bursts occurred sequentially in a super-Active Region NOAA 10486. The average rising rate of the flux density above 200 GHz is only 20 sfu/GHz (corresponding spectral index $\alpha$ of 1.6) for the THz spectral components of 2003 October 28 and November 4 bursts, while it can attain values of 235 sfu/GHz ($\alpha$=4.8) for 2003 November 2 burst. The steeply rising THz spectrum can be produced by a population of  high relativistic electrons with a low-energy cutoff of 1 MeV , while it only requires a low-energy cutoff of 30 keV for the two slowly rising THz bursts, via gyrosynchrotron (GS) radiation based on our numerical simulations of burst spectra in the magnetic dipole field case. The electron density variation is much larger in the THz source than that in microwave (MW) one. It is interesting that the THz source radius decreased by 20--50$\%$ during the decay phase for the three events, but the MW one increased by 28$\%$ for the 2003 November 2 event. In the paper we will present a calculation formula of energy released by ultrarelativistic electrons, accounting the relativistic correction for the first time. We find that the energy released by energetic electrons in the THz source exceeds that in microwave one due to the strong GS radiation loss at THz range, although the modeled THz source area is 3--4 orders smaller than the modeled MW one. The total energies released by energetic electrons via the GS radiation in radio sources are estimated, respectively, to be $5.2\times10^{33}$, $3.9\times10^{33}$ and $3.7\times10^{32}$ erg for the October 28, November 2 and 4 bursts, which are 131, 76 and 4 times as large as the thermal energies of $2.9\times10^{31}$, $2.1\times10^{31}$ and $5.2\times10^{31}$ erg estimated from the soft x-ray GOES observations.  
}

   \authorrunning{A. H. Zhou, J. P. Li \& X. D. Wang }            
   \titlerunning{Diagnostics  From Three Rising Submillimeter Bursts }  

   \maketitle

%
%
\section{Introduction}           
\label{sect:intro}
Solar flares are a consequence of magnetic instabilities in the solar flare regions. During the flares, a large amount of magnetic energy is released into the acceleration of charged particles. A broad spectrum of electromagnetic radiation is emitted. So one of the most direct diagnostic of energetic ($\sim$1 MeV) electrons accelerated during solar flares is their GS radiation at centimeter-millimeter wavelengths emitted in magnetic loops associated with the flaring active region (e.g., Pick et al. 1990; Bastain et al. 1998). Before the year 2000 no radio observations above 90 GHz were available . At such frequencies the characteristic energy of radiating electrons is  of a few MeV (e.g.,  Dulk 1985; Ramaty et al. 1994). Since 2000 new instrumentation observing in the 200--400 GHz range has become available, more than 10 flares have been observed in this band (l{\"u}thi et al. 2004a; l{\"u}thi et al. 2004b; Silva et al. 2007; Krucker et al. 2013).

It is interesting that among of them three strong submillimeter bursts occurred in succession in the same supper-Active Region NOAA 10486 on 2003 October 28, November 2 and 4. For the three events, all the radio spectrum above 200 GHz are not the continuation of the GS spectrum measured at lower frequencies, but surprisingly increases with increasing frequency (l{\"u}thi et al. 2004a; Kaufmann et al. 2004; Silva et al. 2007; Trottet et al. 2008). This spectral feature is termed a "THz component''. The positive-slope THz bursts have been observed thus far in only a handful of the most energetic events (Krucker 2013). So the three THz burst observations occurred in the same active region are very valuable.

The Terahertz wavelength range (0.1-10 THz) is a frontier observational window and its act is not replaced by other wavelength range, because it can provide unique diagnostics about energy release of ultrarelativistic electrons and their environment in lower atmosphere levels from 1000 to 30,000 km above the photosphere in flare regions. The THz events occurred on 2003 November 4 and 2 have been studied briefly ( Zhou et al. 2010; Zhou et al. 2011).  In the paper we will investigate the 2003 Octerbor 28 event in detail . We have carried out a sequence of numerical simulations for the spectral observations, using our GS emission model in the magnetic dipole field case (Zhou et al. 2008). 

In the paper we will present, for the first time, a calculation formula of energy released by energetic electrons in the THz emission region, including the relativistic correction. We will use it to obtain the estimation of the energy  released by energetic electrons in THz and MW emission regions for the three THz events. The total non-thermal energy released by energetic electrons in the radio wavelength range and thermal energy estimated from the soft x-ray GOES observations have been estimated and compared for the three bursts. Finally we  present discussions and conclusions. 


\section{Observations}
\label{sect:Obs}
Extensive flare activities were observed in a super-AR NOAA 10486 during
its disk passage of October 22 -- November 4, 2003. Among them an
extremely energetic 4B/X17.2 flare on October 28, 2003/11:10 was
observed when the active region was located at S16 E08, i.e., close to
the disk-center. The flare was rated the third large X-ray flare recorded by GOES satellite and the largest optical class (4B) flare observed so far. It is associated with a large MW and a rising THz burst. Emission at 210 GHz was first detected by the K{\"o}ln Observatory for Submillimeter and millimeter Astronomy (KOSMA) as a slow rise in intensity at $\sim$11:00 UT (L{\"u}thi et al. 2004a; Trottet et al. 2008), i.e, ten minutes before at the onset of the flare. After a dramatic increase at $\sim$11:02:30 UT, enormous peak flux densities of 25,000 and 11,000 sfu were reached at 11:05:25 UT at 90 and 210 GHz respectively (see Figure 1).

\begin{table}[ht]
	\begin{center}
		\caption{Three Novel Rising THz Burst Observations in the AR 10486.}\label{TAB:tab1}
		\begin{tabular}{ccccccc}
			\hline\noalign{\smallskip}
			Date       & $H_{\alpha}$& $X-ray$      &$ Position$  & $S_{MW}(sfu)     $&$S_{\sim200G}(sfu)$ & $S_{405GHz}(sfu)$ \\
			\hline\noalign{\smallskip}
			10 28 2003/11:02 & $4B $       & $X_{17.2}$   &  $S16E08$   & $S_{90GHz}:25 000  $&$11 000            $ & $      $ \\
			11 02 2003/17:16 & $2B $       & $X_{8.3} $   &  $S18W56$   & $S_{18GHz}:35 000  $&$ 4 000            $ & $70 000$  \\
			11 04 2003/19:42 & $3B $       & $X_{\geq28}$    &  $S19W83$   & $S_{18GHz}:48 000  $&$11 500            $ & $20 000$ \\
			\hline
		\end{tabular}
	\end{center}
\end{table}

The second rising THz burst in AR 10486 was detected by the Solar Submillimeter Telescope (SST) at 212 and 405 GHz (Silva et al. 2007) in the flare on November 2, 2003 starting at $\sim$17:16 UT. This flare is classified as an X8.3 and 2B event. Their peak flux
densities reached values of about 4,000 and 70,000 sfu at 212 and
405 GHz respectively. When the active region passed the west limb of
the solar disk, the third large rising THz burst was observed by SST on November 4, 2003/19:42 UT (Kaufmann et al. 2004). It was
associated with an $X_{\geq28}$ flare (Kane et al. 2005), which may have been the largest X-ray event since observations began in 1976.
The peak flux densities at 18 and 212 GHZ attained, respectively,
values of 48,000 and 11,500 sfu at the maximum phase (see Table 1).

\section{RISING RATE OF FLUX DENSITY OF SUBMILLIMETER SPECTRUM}

\label{sect:incr}
For the rising THz burst on October 28 , emission at 210 GHz was detected as a slow rise in intensity at $\sim$11:00 UT. The total flux density time profile exhibits a slowly varying, time-extended component from an extended source and a short-lived component from a compact source exhibiting three distinctive peaks. However there are no significant differences between the spectra emitted by the extended and the compact sources (L{\"u}thi et al.
2004a). The flux density at 210 GHz increased from 3,100 to 11,000
sfu in a period of 11:03:15 to 11:05:25 UT at the rise phase, but the 230 and 345 GHz KOSMA-channels were saturated at these times due to the enormous flux density of the burst. So the corresponding flux densities have not been recorded at 230/345 GHz during the main phase. Figure 1 shows the temporal evolution spectra of this event given by L{\"u}thi et al. (2004a).

The rise rate \emph{r} of the flux density measured from the observation spectra is in the 18.5-8.5 sfu/GHz range during the October 28 burst (see Table 2), i.e., it is a slowly rising THz burst. The second THz burst occurred on November 4, exhibiting four peaks. Its rising rates of the flux density are also given in Table 2. It shows that for the 2003 November 4 event, \emph{r} value variation is in the range of 7.8--44 sfu/GHz, which means that it is also a slowly rising THz burst. Their average rising rate reaches only 20 sfu/GHz (corresponding spectral index $\alpha$ of 1.6) for the two events. The rising rates of a steeply rising THz burst of the 2003 November 2 event were estimated (Li et al. 2015). Its average rise  rate  can attain a value of 235 sfu/GHz ($\alpha$=4.8) for 2003 November 2 burst,  which is about one order of magnitude higher than that for the two slowly rising THz burst. 

\begin{figure}
	\centering
	\includegraphics[width=0.8\textwidth, angle=0]{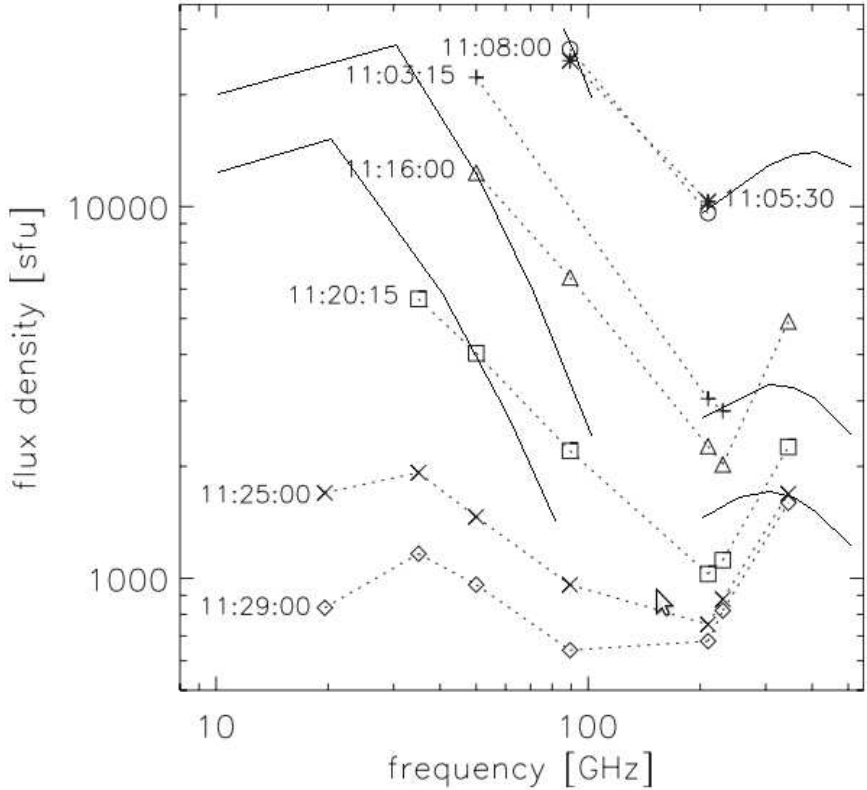}
	\caption{The temporal evolutions of radio spectra of the October 28
		burst given by L{\"u}thi et al. (2004a) and for their fits (see the solid lines).}
	\label{Fig:fig1}
\end{figure}

\begin{table}[ht]
	\begin{center}
		\caption{Rising Rates r (sfu/GHz) of the Flux Density  of the THz component at the Rise, Maximum Phase and Decay Phase
			for the three THz Bursts, measured from the observations at 210, 230/345 GHz (KOSMA) and at 212 and 405 GHz (SST).}\label{TAB:tab2}
		\begin{tabular}{ccccccccc}
			\hline\noalign{\smallskip}
			date      &$  time  $&$  Rise-phase      $&$ Max.-phase    $&$Decay-phase $&$S_{\sim200GHz} $&$S_{345 or 405GHz}$&$r~sfu/GHz$\\
			\hline\noalign{\smallskip}
			2003 10 28&$11:03:15$&$  yes          $&$            $&$            $&$3.1\times10^{3}$&$                $&$     $\\
			&$11:05:30$&$               $&$ yes        $&$            $&$1.1\times10^{4}$&$                $&$     $\\
			&$11:16:00$&$               $&$            $&$ yes        $&$2.2\times10^{3}$&$ 4.7\times10^{3}$&$18.5$\\
			&$11:20:15$&$               $&$            $&$ yes        $&$1.2\times10^{3}$&$ 2.3\times10^{3}$&$8.5$\\
			&$11:25:00$&$               $&$            $&$ yes        $&$8.5\times10^{2}$&$ 2.7\times10^{3}$&$13.7$\\
			&$11:29:00$&$               $&$            $&$ yes        $&$8.0\times10^{2}$&$ 2.6\times10^{3}$&$13.3$\\
			2003 11 02&$17:16:15$&$  yes          $&$            $&$            $&$1.2\times10^{3}$&$ 3.1\times10^{4}$&$154$\\
			&$17:17:06$&$               $&$ yes        $&$            $&$4.0\times10^{3}$&$ 7.0\times10^{4}$&$342$\\
			&$17:17:30$&$               $&$            $&$ yes        $&$3.2\times10^{3}$&$ 5.0\times10^{4}$&$242$\\
			&$17:18:00$&$               $&$            $&$ yes        $&$3.5\times10^{3}$&$ 4.0\times10^{4}$&$210$\\
			&$17:18:30$&$               $&$            $&$ yes        $&$4.0\times10^{3}$&$ 5.8\times10^{4}$&$280$\\
			&$17:19:00$&$               $&$            $&$ yes        $&$5.0\times10^{3}$&$ 5.5\times10^{4}$&$259$\\
			&$17:19:30$&$               $&$            $&$ yes        $&$5.0\times10^{3}$&$ 5.5\times10^{4}$&$259$\\
			&$17:20:00$&$               $&$            $&$ yes        $&$5.0\times10^{3}$&$ 4.8\times10^{4}$&$223$\\
			&$17:21:00$&$               $&$            $&$ yes        $&$4.5\times10^{3}$&$ 3.2\times10^{4}$&$142$\\
			2003 11 04&$19:42:40$&$  yes          $&$            $&$            $&$2\times10^{3}  $&$   5\times10^{3}$&$15.5$\\
			Peak 1   &$19:44:05$&$               $&$ yes        $&$            $&$1.15\times10^{4}$&$2.0\times10^{4}$&$44$\\
			Peak 2   &$19:45:20$&$               $&$            $&$ yes        $&$          10^{4}$&$1.65\times10^{4}$&$33.7$\\
			Peak 3   &$19:46:50$&$               $&$            $&$ yes        $&$          10^{4}$&$1.5\times10^{4}$&$25.9$\\
			Peak 4   &$19:48:25$&$               $&$            $&$ yes        $&$3.7\times10^{3} $&$ 5.2\times10^{3}$&$7.8$\\
			
			\hline
		\end{tabular}
	\end{center}
\end{table}

\section{FITS OF THE THREE RISING SUBMILLIMETER BURST SPECTRA}
\label{sect:fits} It is well known that radio spectrum can provide
crucial information about energetic electrons and their environment
in solar flares. The information contains mainly the energy spectral
index $\delta$, low-energy and high-energy cutoffs $E_{0}$ and
$E_{m}$, electron number density $\emph{N}$, source size, and magnetic field strength $\emph{B}$ in source region. 

\subsection{For the two slowly rising THz bursts}
Here we will model the slowly rising THz spectral
components of the 2003 October 28 burst for the first time. For this rising
THz burst, the flux density at 210 GHz increased from
3,100 to 11,000 sfu at the rise phase but the corresponding higher
frequency ($\nu>210~GHz$) observations have not been obtained
during the main phase. So we only can estimate what condition, at
least, can produce the rising THz spectral component
with 11,000 sfu flux density at 210 GHz at the maximum phase via the GS
emission, which leads to that the modeled spectrum is underestimated 
largely. It is well known that the low-energy cutoff and electron
density can affect substantially the spectral calculations, so we
selected, respectively, a sequence of low-energy cutoffs $E_{0}$ and
a sequence of electron number densities $\emph{N}$ to model
THz burst spectra for $E_{m}=10~ MeV$. We find from these spectral calculations that the best set of values for the THz burst spectral fit at 11:05:30 UT of the maximum phase are for the low-energy cutoff of 30 keV and the number density of $4.5\times10^{10}$ $cm^{-3}$, where $\delta$=2,  $B_{0}=5000$ G, $\theta=10^{\circ}$, and $h_{d}=10^{8}$ $cm$. The another two THz spectra at 11:16:00 and 11:20:15 UT at the decay phase also are fitted. The modeled THz and MW emission spectra are given in Figure 1 in the case of magnetic dipole field , which are superimposed on the original Figure 10 (dotted lines) given by L\"{u}thi et al. (2004a). It is shown that the modeled spectra fit well with the observational ones of the October 28 burst at 11:05:30, 11;08:00, 11:16:00, and 11:20:15 UT (see solid lines). The physical parameters used in the spectral calculations are given in Table 3. We can find from Table 3 that the low-energy cutoff were nearly constant during the THz burst, but the required number density of electrons decreased substantially from $4.5\times10^{10}$ to $4.5\times10^{8}$ $cm^{-3}$ at the decay phase. The fit results for the MW spectra of the October 28 burst also be given in Table 3 during the burst for $E_{0}=10~ keV$ and $E_{m}=5~ MeV$. At the decay phase the electron number density $\emph{N}$ in the MW source decreased from $6\times10^{7}$ to $1.5\times10^{6}$ $cm^{-3}$, i.e., decreased by 40 times, but the value of $\emph{N}$  decreased by 100 times in the THz source. The total electron number $N_{total}$ decreased by $\sim$40 and $\sim$400 times in the MW and THz source, respectively.

\begin{table*}
	\caption{Physical Parameters of Energetic Electrons for the Three Bursts}\label{TAB:tab3}
	\begin{tabular}{ccccccccc}
		\hline\noalign{\smallskip}
		date&
		Time &$\delta$& MW:~$R\arcsec$& ${N~cm^{-3}}$&${N_{total}}$& $THz:~R\arcsec$&${N~cm^{-3}}$& ${N_{total}}$\\
		\hline\noalign{\smallskip}
		2003 10 28&
		11: 05 : 30&$2$&$25 $  &$6.0\times10^{7}$&$2.0\times10^{35}$&$0.5    $&$4.5\times10^{10}$&$5.9\times10^{34}$\\
		&$11:16:00$&$2.2$&$25 $  &$2.0\times10^{7}$&$6.6\times10^{34}$&$0.35    $&$6\times10^{9}$&$3.8\times10^{33}$\\
		&$11:20:15$&$1.9$&$25 $  &$1.5\times10^{6}$&$4.9\times10^{33}$&$0.25    $&$4.5\times10^{8}$&$1.5\times10^{32}$\\
			
		2003 11 2&
		17 : 16 : 15&$3$&$25 $  &$8.0\times10^{7}$&$2.6\times10^{35}$&$0.5    $&$8\times10^{6}$&$1.0\times10^{31}$\\
		&$17:17:06$&$3$&$25   $&$1.8\times10^{8}$&$5.9\times10^{35}$&$0.5    $&$4\times10^{8}$&$5.2\times10^{32}$\\
		&$17:17:30$&$3$&$25   $&$1.6\times10^{8}$&$5.3\times10^{35}$&$0.5    $&$       10^{8}    $&$1.3\times10^{32}$\\
		&$17:18:00$&$3$&$25   $&$1.6\times10^{8}$&$5.3\times10^{35}$&$0.5    $&$4\times10^{7}$&$5.2\times10^{31}$\\
		&$17:18:30$&$3$&$25   $&$1.6\times10^{8}$&$5.3\times10^{35}$&$0.5    $&$3\times10^{8}$&$3.9\times10^{32}$\\
		&$17:19:00$&$3$&$25   $&$1.5\times10^{8}$&$5.0\times10^{35}$&$0.5    $&$2\times10^{8}$&$2.6\times10^{32}$\\
		&$17:19:30$&$3$&$30   $&$1.3\times10^{8}$&$6.1\times10^{35}$&$0.45   $&$2\times10^{8}$&$2.2\times10^{32}$\\
		&$17:20:00$&$3$&$30   $&$1.3\times10^{8}$&$6.1\times10^{35}$&$0.45   $&$1.3\times10^{8}$&$1.4\times10^{32}$\\
		&$17:21:00$&$3$&$32   $&$1.3\times10^{8}$&$7.0\times10^{35}$&$0.38   $&$7\times10^{7}$&$5.3\times10^{31}$\\
		2003 11 4&
		P1&$2.3$&$40 $  &$1.2\times10^{6}$&$1.0\times10^{34}$&$0.5    $&$1.0\times10^{10}$&$1.0\times10^{34}$\\
		&P2&$2.3$&$40 $  &$6.0\times10^{5}$&$5.0\times10^{33}$&$0.25    $&$5.5\times10^{9}$&$1.8\times10^{33}$\\
		
		\hline
	\end{tabular}
\end{table*}

Another slowly rising THz burst on November 4 associated the largest soft X-ray burst ($X_{\geq28}$) so far. However the associated rising THz spectral components are not so strong and the rising rates are only in the 7.8--44 rang (see Table 1), so it also belongs to a slowly rising THz burst. Their spectral fit results of the peak 1 and peak 4 had been given (see the original Figure 2 and Table 2 given by Zhou et al. 2011). The required the high-energy cutoff is also only 30 keV as the 2003 October 28 THz burst. The flux density reaches 11,500 sfu at 212 GHz at peak 1, which is close to the peak flux density of 11,000 sfu at 210 GHz of the October 28 burst, but the required electron number density for the November 4 burst is only $10^{10}$ $cm^{-3}$ (see Table 3), which is only $\sim$ 1/5 of the required value ($4.5\times10^{10}$ $cm^{-3}$) of the October 28 burst. In the decay phase the $\emph{N}$ and ${N_{total}}$ values decreased about one and 5 times in the THz source, respectively.

\subsection{For the steeply rising THz burst}
A giant rising THz burst detected on 2003 November 2 in the Super-AR NOAA 10486. Observations show that the flux density of the THz spectrum steeply rising  and their rising rate of the flux density of the THz spectrum
attained as high as 342 sfu/GHz at the maximum phase. Their mean rising rate also reached  a value of 235 sfu/GHz (corresponding spectral index $\alpha$ of 4.8) during the burst (Li et al. 2015). The flux densities reached about 4,000 and 70,000 sfu at 212 and 405 GHz at the maximum phase respectively. The emissions at 405 GHz maintained continuous high level that they  exceed largely the peak values of the microwave (MW) spectra during the main phase. Our studies suggest that such  strong and steeply rising THz component can be produced by energetic electrons with a low-energy cutoff of 1 MeV via GS radiation in the magnetic dipole field case (Li et al. 2015). The electron number density $\emph{N}$ derived from  our numerical fits, increased substantially from $8\times10^{6}$  to $4\times10^{8}$ cm$^{-3}$ at the rise phase, i.e., $\emph{N}$ value increased 50 times at the rise phase  (see Table 3).  During the decay phase it decreased to $7\times10^{7}$ cm$^{-3}$, i.e., decreased about five times from the maximum phase. The total electron number decreased an order of magnitude  at the decay phase. Nevertheless in the MW emission source the $\emph{N}$ value decreased only by $\sim 30 \%$ and the total electron number did not decrease but increased by $\sim20\%$ at the decay phase. 

The fit parameters at the maximum phase for the three radio events
are given in Table 4. It is found from it  that the required
electron number density reaches $\sim10^{10}~cm^{-3}$ for the two
slowly rising THz burst spectra at the maximum phase, which is
two orders of magnitude higher than that for the steeply increasing one.
But for the steeply rising THz spectrum it requires  a much
higher low-energy cutoff of 1 MeV, while for  the two slowly
rising THz burst spectrum it only requires a 30 keV low-energy
cutoff. 

\section{THE ENERGY FLUX OF ENERGETIC ELECTRONS}
\label{sect:energy} The energy flux and energy released by energetic
electrons are important constraints on acceleration
mechanisms (Miller et a. 1997). These quantities are sensitive to
the low-energy cutoff in the electron distribution (Holman, 2003 ).
Once the energy cutoffs and the number density of the energetic
electrons are obtained from the numerical fit of observational
spectrum, the distribution function of energetic electron n(E)=G$E^{-\delta}$ and the instantaneous energy flux $E_{F}$ carried by energetic electrons can be determined as well. Here we will present a calculation formula of $E_{F}$  for any time, including the relativistic correction factor $\gamma$ (Lorentz factor) (c.f. Zhou et al. 1996; 2011)  for the first time. It is 

\begin{eqnarray}
E_{F}\simeq\frac{3.0G}{2.5-\delta}(E_{m}^{2.5-\delta}-E_{0}^{2.5-\delta})\gamma~~~~(\delta\neq2.5), \nonumber\\
E_{F}\simeq{3.0G}ln(E_m/E_0)\gamma~~~~  (erg~cm^{-2}~s^{-1})     (\delta= 2.5).
\end {eqnarray}

 The G factor is
\begin{eqnarray}
	G=\frac{N(\delta-1)}{(E_{0}^{1-\delta}-E_{m}^{1-\delta})}~~~(\delta\neq1), \nonumber\\
	G=N/ln(E_m/E_0)~~~	(\delta=1).
	\end {eqnarray}

Lorentz factor $\gamma$ is a function of electron energy. Here they are taken as 2 and 7.3 corresponding respectively to 500 keV and 3.2 MeV  for the two slowly and a steeply rising THz bursts. Then we can estimate the instantaneous energy flux $E_{F}$ at the maximum time putting these electron parameters (see Table 4) into the equations (1) and (2). The energy loss rate from the GS radiation, $E'~erg~ s^{-1}$(=$E_{F}\times$A) can be estimated on the the source area A. Finally the energy E erg(=$E'\times\Delta~T)$ released by energetic electrons via the GS radiation can also be calculated on the lifetime $\Delta$T (second) (full width at half
maximum for the burst time profile). The estimated energy flux,
energy loss rate, and energy released by energetic electrons are
given in Table 5  in the THz and MW sources for the three bursts on the physical parameters given in Table 4.

\begin{table}[ht]
\begin{center}
\caption{Parameters of the Burst Sources and Energetic Electrons at
	the Maximum Phase for the Three Rising THz 	Bursts.}\label{TAB:tab4}
\begin{tabular}{ccccccccc}
\hline\noalign{\smallskip}
date             & $B_{0}(G)$&$\theta^\circ$&$R\arcsec$&$\delta$&$E_{0}(keV)$&$E_{m}(MeV)$&$N(cm^{-3})$&$N_{total}$\\
\hline\noalign{\smallskip}
2003 10 28 (THz) & $ 5000  $ & $  10     $ & $ 0.5  $ & $2.0 $ & $  30    $& $10 $& $4.5\times10^{10}$&$5.9\times10^{34}$\\
11 02       & $ 5000  $ & $  60     $ & $ 0.5  $ & $3.0 $ & $  1000  $& $10 $& $4\times10^{8}   $&$5.2\times10^{32}$\\
11 04       & $ 5000  $ & $  80     $ & $ 0.5  $ & $2.3 $ & $  30    $& $10 $& $      10^{10}   $&$1.3\times10^{34}$\\
2003 10 28 (MW)  & $ 2800  $ & $  10     $ & $  25  $ & $2.0 $ & $  10    $& $ 5 $& $6\times10^{7}   $&$2.0\times10^{35}$\\
11 02       & $ 2800  $ & $  60     $ & $  25  $ & $3.0 $ & $  10    $& $ 5 $& $1.8\times10^{8} $&$5.9\times10^{35}$\\
11 04       & $ 2000  $ & $  80     $ & $  40  $ & $2.3 $ & $  10    $& $ 5 $& $1.2\times10^{6} $&$1.0\times10^{34}$\\
\hline
\end{tabular}
\end{center}
\end{table}

Table 5 shows that the energy flux $E_{F}$ carried by the energetic
electrons reached $1.5\times10^{15}$, $8\times10^{14}$, and
$1.4\times10^{14}$ erg cm$^{-2}$ s$^{-1}$ at the maximum phase in the THz source for the three bursts, respectively. However in the MW source they only reached $2.4\times10^{11}$, $6.6\times10^{10}$, and $1.8\times10^{9}$  erg cm$^{-2}$ s$^{-1}$ respectively, which are 3-5 orders of magnitude lower than that in the THz source. The energy loss rate $E'$ reached, respectively,  $6.1\times10^{30}$ -- $5.7\times10^{29}$ and
$4.8\times10^{28}$ -- $2.5\times10^{30}~erg~s^{-1}$ ranges in the
THz source and in MW one at the maximum phase. It is found from Table 5 that although the modeled submillimeter source area is 3-4 orders of magnitude smaller than the modeled MW one, while the energy ($E_{THz}$) released by energetic electrons in the THz emission source exceeds that ($E_{MW}$) in microwave one. The ratio of $E_{THz}$ to $E_{MW}$ is 2.4, 5.0, and 12 for the three events, respectively. The total energy $E_{R}$ released by energetic electrons in THz and MW sources reached $3.8\times 10^{33}$, $1.6\times 10^{33}$, and $1.8\times 10^{32}$ erg for the October 28 burst, November 2 and 4 bursts, respectively (see Table 6). So in view of the radio energy the October 28 burst is the strongest for the three events.

\begin{table}[ht]
\begin{center}
\caption{Energy Flux $E_{F}$, Energy Loss Rate $E'$, and Total
	Energy E Carried by Energetic Electrons vis the GS Radiation in the
	THz and MW Sources for the Three Submillimeter
	Bursts.}\label{TAB:tab5}
\begin{tabular}{ccccccc}
\hline\noalign{\smallskip}
$date         $ &$ N(cm^{-3})     $&$E_F(erg~cm^{-2}s^{-1})$&$E'(erg~s^{-1})  $&$\Delta~T(s)$&$ E(erg)         $ \\
\hline\noalign{\smallskip}
2003 10 28(THz) &$4.5\times10^{10}$&$1.5\times10^{15}      $&$6.1\times10^{30}$&$        450     $&$2.7\times10^{33}$\\
11 02                  &$4.0\times10^{8}   $&$8.0\times10^{14}      $&$3.3\times10^{30}$&$        380$&$1.3\times10^{33}$\\
11 04                  &$10^{10}         $&$1.4\times10^{14}      $&$5.7\times10^{29}$&$        300$&$1.7\times10^{32}$\\
10 28 (MW) &$6\times10^{7}   $&$2.4\times10^{11}      $&$2.5\times10^{30}$&$        450$&$1.1\times10^{33}$\\
11 02      &$1.8\times10^{8} $&$6.6\times10^{10}      $&$6.9\times10^{29}$&$        380$&$2.6\times10^{32}$\\
11 04      &$1.2\times10^{6} $&$1.8\times10^{9}       $&$4.8\times10^{28}$&$        300$&$1.4\times10^{31}$\\
\hline
\end{tabular}
\end{center}
\end{table}

\section{Discussion}
\label{sect:discussion}
\subsection{Propagation effect} 
Although flux density at 210 GHz of the October
28 burst is smaller than the value at 212 GHz of the November 4
burst, the required electron number density reaches as high as
$4.5\times10^{10}~cm^{-3}$, which is 3.5 times higher than that of
the November 4 burst maybe due to propagation effect. It was found that
the emissivity of GS radiation increases with the propagation angle for the same harmonic number in the MW and millimeter range.  And the increasing trend becomes more obviously (Zhou et al. 1999). In the THz range the propagation effect can be clarified by Figure 2. It shows the different GS emission spectra in the case of different propagation angles
$\theta$, where $\delta=3$, $E_{0}=500~keV$, $E_{m}=10~MeV$, and $N=10^{7}~cm^{-3}$. We can see from it that the flux densities at the higher frequencies in the THz range for $\theta=80^\circ$ are, at least, one order of
magnitude higher than those for $\theta=20^\circ$ . The propagation
effect results in a higher electron number density requirement under
the quasi-longitudinal propagation than that under the
quasi-transverse one for a same observational flux density distribution.

\begin{figure}
\centering
\includegraphics[width=0.8\textwidth, angle=0]{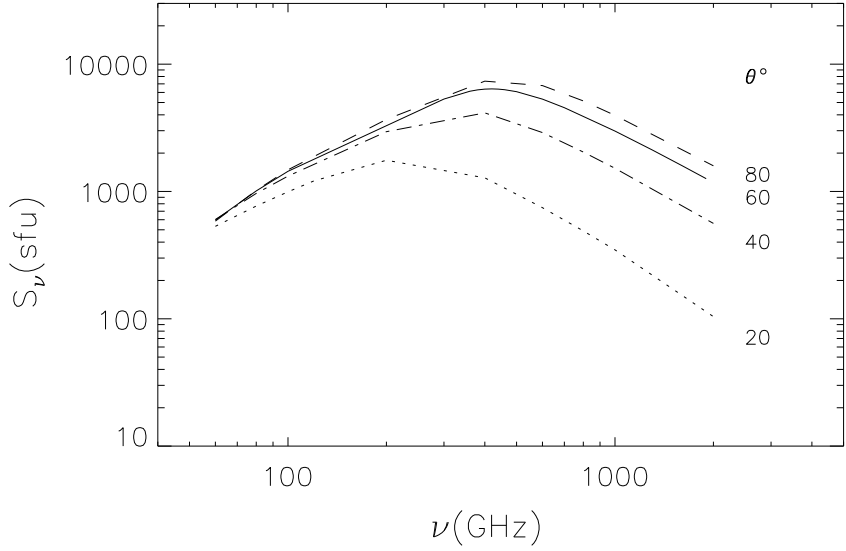}
\caption{The effect of the
	propagation angles $\theta^\circ$ on  the GS emission spectrum in
	submillimeter range,  where $\delta=3$, $E_{0}=500~keV$, $E_{m}=10~MeV$, $N=10^{7}~cm^{-3}$, and $B_{0}=5000~G$. }
\label{Fig:fig2}
\end{figure}

\subsection{Source size variation}
We find that all the flux densities decreased rapidly in the THz range at decay phase for the three THz bursts. If we still take the same source size
($R=0.\arcsec5$), then the required electron number $\emph{N}$ will
decrease largely, which lead to the modeled flux densities of the GS
emission at 345 or 405 GHz being  always lower than the observational
results, i.e., the modeled rising rate is smaller than that the
observational one. So that the modeled GS spectrum can not fit the
observational spectrum at higher frequencies. In this case we have
to take a smaller source size of $0.\arcsec38$, $0.\arcsec35$ , and even to $0.\arcsec25$ to fit these spectra in the decay phase for the three THz bursts, i.e., the  THz source radius decreased by 20-50 $\%$ in the decay phase. The effect of the emission source size on the GS emission spectrum in the THz range is given (Li et al. 2015). On the contrary, we also found that the MW source size obtained from the spectrum fit increased from 25$\arcsec$ to 32$\arcsec$ at the decay phase of the November 2 burst, i.e., the MW source radius increased by 28  $\%$ . This source size variation maybe is a rather interesting result. It would be a reflect of the various changes from the energetic electron acceleration,  trapping, and the magnetic field topology in burst source.

\subsection{Comparison of radio energy and thermal energy}
The energy $E_{THz}$ and $E_{MW}$ released by energetic electrons of
the October 28 burst in the THz and MW ranges can attain values of $2.7\times10^{33}$ and  $1.1\times10^{33}~erg$, respectively, which are
the highest for the three bursts. The total radio energy $E_{R}$ in the
THz and MW ranges of the burst can reach $3.8\times10^{33}~ erg$ due to a hard electron spectral index of 2 and a high electron number density of
$4.5\times10^{10}~cm^{-3}$ (see Table 4). So in view of radio energy the October 28 burst is the strongest for the three events. The ratio of the radio
energy to the thermal energy,  $E_{R}/E_{T}$ is 131 times for the
October 28 burst, i.e., the radio energy is two orders of
magnitude higher than that the thermal energy estimated from the
soft X-ray GOES observations for the emission measure and
temperature. For the November 2 burst $E_{THz}$ reached only $1.3\times10^{33}~erg$ due to a narrower energy release range of
electrons from 1 to 10 MeV and a mean electron number density. The
value of $E_{R}/E_{T}$ is 76 for this burst. For the November 4
burst the $E_{R}/E_{T}$ value is only 4, because it is
associated with the largest soft X-ray flare so far and the estimated
thermal energy attained $5.2\times10^{31}~erg$.

\subsection{Comparison of the modeled spectra of the three radio bursts}
Figure 3 shows a comparison of the three modeled GS spectra fitting
the observations of the October 28, November 2 and 4 bursts over
interval of the maximum phase or at the maximum phase. It shows that the MW emission of the October 28 burst is the strongest for the three bursts, because it is produced by the energetic electrons with a harder spectral index ($\delta=2$). While the THz emission of the October 28 burst appears to be lower , although the observational flux density at 210 GHz is close to that of the another two bursts. It results from that the corresponding higher frequency ($\nu>210~GHz$) observations  have not been obtained during the main phase. So we only give an increasing THz spectrum component with a smaller rising rate, which leads to an underestimated model spectrum. For the November 2 burst the modeled THz emission is the strongest among the three bursts and this peak frequency reaches 1440 GHz, due to intense ultra-relativistic electron GS radiation  in a higher energy emission range of 1-10 MeV under the quasi-transverse propagation condition. It is shown from the comparison of the modeled spectra of the three THz bursts
that the emission strengths are very different for the three bursts
and for the different emission frequency ranges, which depends
strongly on the electron acceleration and various physical conditions of
burst region.

\begin{figure}
	\centering
	\includegraphics[width=0.8\textwidth, angle=0]{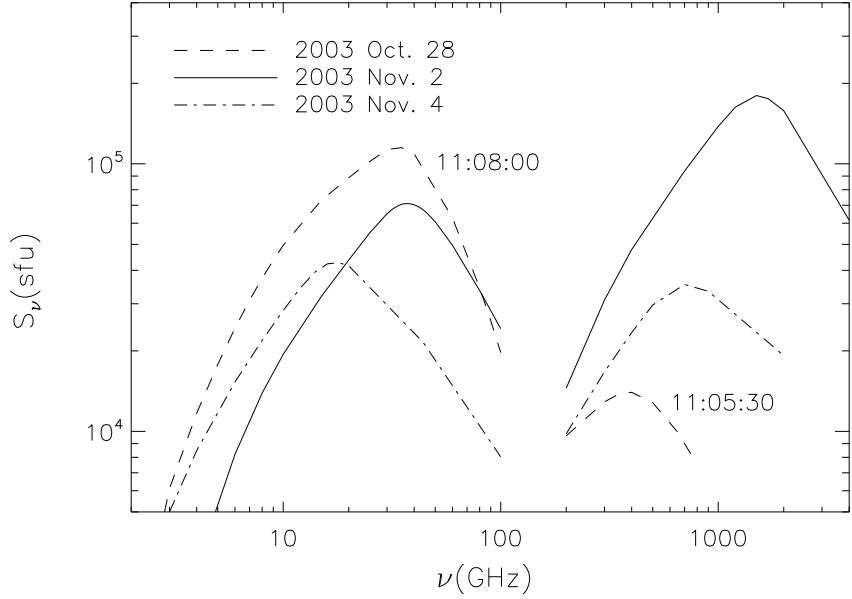}
	\caption{The modeled spectra for the
		observations at 11:08:00 and 11:05:30 UT over interval of the
		maximum phase of the 2003 October 28 burst, at 17:17:06 UT of the
		maximum phase of the November 2 burst, and at the peak 1 of the
		maximum phase of the November 4 burst (Zhou et al.
		2011).}
	\label{Fig:fig3}
\end{figure}

	\begin{table}[ht]
	\begin{center}
	\caption{Total Radio Energy $E_{R}(erg)$ Carried by Energetic
		Electrons and the Thermal Energy $E_{T}(erg)$ Estimated from the
		Soft X-Ray GOES Observations.} \label{TAB:tab6}
	\begin{tabular}{cccccccc}
	\hline\noalign{\smallskip}
	$date       $ &$E_{THz}(erg)  $ & $E_{MW}(erg)     $ & $E_{THz}  /E_{MW} $&$E_{R}(erg)$ & $E_{T}(erg)          $ &$E_{R}/E_{T} $\\
	\hline\noalign{\smallskip}
	2003 10 28 &$2.7\times10^{33}$&$1.1\times10^{33}$&$2.4$&$3.8\times10^{33}$&$2.9\times10^{31}$&$131        $\\
	11 02 &$1.3\times10^{33}$&$2.6\times10^{32}$&$5.0$&$1.6\times10^{33}$&$2.1\times10^{31}$&$76        $\\
	11 04 &$1.7\times10^{32}$&$1.4\times10^{31}$&$12$&$1.8\times10^{32}$&$5.2\times10^{31}$&$ 4        $\\
	\hline
	\end{tabular}
	\end{center}
	\end{table}
	
\section{Conclusions}
\label{sect:conclusion} In the paper we investigate the three novel
rising submillimeter bursts occurring in the Active Region NOAA 10486. It is found from the numerical fit that the two slowly rising  and a steeply rising submillimeter spectral components can be produced, respectively, by energetic electrons with 30keV--1 MeV and 1--10 MeV energy ranges  in a compact source (about 0.$\arcsec5$ radius) with strong local magnetic fields varying from 4590 to 780 G via the GS emission. The photosphere magnetic field of 5000 G would be possible on observation in a compact source (Li et al. 2015). The associated microwave spectral components can be produced by energetic electrons with 10 keV-- 5MeV and  with a mean local magnetic field strength in an extend source of 25$\arcsec$--40$\arcsec$ radius. 

It is found from the spectral temporal evolution that the number
density variation amplitude is much larger in the THz
source than that in the MW source during the bursts. The dramatic
variation of electron number density in the THz source
could result from the effective electron acceleration in the rise
phase and strong electron energy loss during the flare. While in the
MW source the variation amplitude of electron number density is one
order of magnitude lower than that in the THz source. Because in the MW source  there are much more electrons decayed from the higher energy to lower energy and less electron energy loss. The
instantaneous energy flux of electrons in the THz source is about 4--5
orders of magnitude higher than that in microwave one for the
three bursts. Although the modeled THz source area is 3--4
orders of magnitude smaller than the modeled  MW one, the energies
released by energetic electrons in the THz source are 2--12 times of those in microwave source due to the strong
GS radiation loss at the submillimeter wavelengths. The total energies
released by energetic electrons via the GS radiation in the MW and
THz sources are estimated, respectively, to be
$3.8\times10^{33}$, $1.6\times10^{33}$ and $1.8\times10^{32}$ erg
for the October 28, November 2 and 4 bursts, which are 131, 76, and 4
times as large as the thermal energies of $2.9\times10^{31}$,
$2.1\times10^{31}$ and $5.2\times10^{31}$ erg estimated from the
soft X-ray GOES observations.

Our investigations show that the detailed GS emission models fit well with  the rising submillimeter spectral components for the three novel submillimeter
bursts. So this submillimeter spectral component could provide important diagnostics about the high relativistic electrons with higher energy
rang of a few tens keV--$\sim$10 MeV and their environment in burst regions. Further more, it is found from
the modeled calculations that the THz source radius decreased by 20--50$\%$ during the decay phase for the three events, but the MW one increased by 28$\%$ for the 2003 November 2 event. The interesting result about source size variations maybe is significant for the studies about energetic
electron acceleration, trapping, and magnetic field construction variation of
source region. However we must note that the required source radius
is usually much smaller, based on the GS emission calculations.
Further progress in understanding the physics of THz emission from
flares requires more observations with a complete spectral coverage
at the THz range, such as the Atacama Large Millimeter/Submillimeter Array (ALMA).

\begin{acknowledgements}
	The authors thank  Dr. V. F. Melnikov for fruitful discussions. This
	study is supported by the NFSC project with Nos.11333009 and 11573072, the ``973'' program with No. 2014CB744200.
\end{acknowledgements}

\label{lastpage}

\end{document}